\documentclass[prl,floatfix,showpacs,superscriptaddress,amsmath,twocolumn]{revtex4}
\usepackage{amssymb}
\usepackage{graphics}
\usepackage{epsfig}
\usepackage{graphicx}

\usepackage{dcolumn}
\usepackage{amsmath,amssymb}
\usepackage{bm}
\usepackage[dvips]{color}

\begin{document}

\definecolor{mygrey}{gray}{0.35}
\definecolor{mygreen}{rgb}{0.85,1,0.9}
\definecolor{myzard}{cmyk}{0,0,0.05,0}
\definecolor{mywhite}{rgb}{1,1,1}
\definecolor{myred}{rgb}{1,0,0}

\def\C{{\mathbb{C}}} \def\F{{\mathbb{F}}}
\def\N{{\mathbb{N}}} \def\Q{{\mathbb{Q}}}
\def\R{{\mathbb{R}}} \def\Z{{\mathbb{Z}}}

\def\cH{{\mathcal H}}

\def\bfsigma{\boldsymbol{\sigma}}
\def\bfmu{\boldsymbol{\mu}}

\def\dd{\mathord{\rm d}} \def\Det{\mathop{\rm Det}}
\def\dist{\mathop{\rm dist}} \def\ee{\mathord{\rm e}}
\def\End{\mathord{\rm End}} \def\ev{\mathord{\rm ev}}
\def\id{\mathord{\rm id}} \def\ii{\mathord{\rm i}}
\def\min{\mathord{\rm min}} \def\mod{\mathord{\rm mod}}
\def\prob{\mathord{\rm prob}} \def\tr{\mathop{\rm Tr}}
\def\half{\textstyle\frac{1}{2}} \def\third{\textstyle\frac{1}{3}}
\def\fourth{\textstyle\frac{1}{4}}

\def\vec#1{{\bf{#1}}} \def\vect#1{\vec{#1}}

\def\bra#1{\langle#1|} \def\ket#1{|#1\rangle}
\def\braket#1#2{\langle#1|#2\rangle}
\def\bracket#1#2{\langle#1,#2\rangle} \def\ve#1{\langle#1\rangle}

\def\lR{l^2_{\mathbb{R}}}
\def\RR{\mathbb{R}}
\def\E{\mathbf e}
\def\D{\boldsymbol \delta}
\def\S{{\cal S}}
\def\T{{\cal T}}
\def\dd{\delta}
\def\one{{\bf 1}}
\def\ss{\boldsymbol \sigma}

\newtheorem{theorem}{Theorem}
\newtheorem{lemma}{Lemma}
\newtheorem{definition}{Definition}

\title{Relativity and Lorentz Invariance of Entanglement Distillability}

\author{L. Lamata}

\affiliation{Instituto de Matem\'aticas y F\'{\i}sica Fundamental,
CSIC, Serrano 113-bis, 28006 Madrid, Spain}

\author{M. A. Martin-Delgado}

\affiliation{Departamento de F\'{\i}sica Te\'orica, Universidad
Complutense de Madrid, E-40036, Spain}

\author{ E. Solano}

\affiliation{Physics Department, ASC, and CeNS,
Ludwig-Maximilians-Universit\"at, Theresienstrasse 37, 80333 Munich,
Germany}

\affiliation{Max-Planck-Institut f\"ur Quantenoptik, Hans-Kopfermann-Str. 1, D-85748 Garching,
Germany}

\affiliation{Secci\'on F\'{\i}sica, Departamento de
Ciencias, Pontificia Universidad Cat\'olica del Per\'u, Apartado
Postal 1761, Lima, Peru}


\begin{abstract}
We study entanglement distillability of bipartite mixed spin states
under Wigner rotations induced by Lorentz transformations. We define
weak and strong criteria for relativistic {\it isoentangled} and
{\it isodistillable} states to characterize relative and invariant
behavior of entanglement and distillability. We exemplify these
criteria in the context of Werner states, where fully analytical
methods can be achieved and all relevant cases presented.
\end{abstract}

\maketitle

Entanglement is a quantum property that played a fundamental role
in the debate on completeness of quantum mechanics. Nowadays,
entanglement is considered a basic resource in present and future
applications of quantum information, communication, and
technology~\cite{NielsenChuang,rmp}. However, entangled states are
fragile, and interactions with the environment destroy their
coherence, thus degrading this precious resource. Fortunately,
entanglement can still be recovered from a certain class of states
which share the property of being distillable. This means that
even in a decoherence scenario, entanglement can be extracted
through purification processes that restore their quantum
correlations~\cite{Bennett1,generaldistill}. An entangled state
can be defined as a quantum state that is not separable, and a
separable state can always be expressed as a convex sum of product
density operators~\cite{Werner}. In particular, a bipartite
separable state can be written as $\rho=\sum_iC_i\rho^{(\rm
a)}_i\otimes\rho^{(\rm b)}_i$, where $C_i\geq0$, $\sum_i C_i=1$,
and $\rho^{(\rm a)}_i$ and $\rho^{(\rm b)}_i$ are density
operators associated to subsystems $\rm A$ and $\rm B$.

In quantum field theory, special relativity
(SR)~\cite{Rindler,VerchWerner} and quantum mechanics are
described in a unified manner. From a fundamental point of view,
in addition, it is relevant to study the implications of SR on the
modern quantum information theory (QIT)~\cite{SRQIT}. Recently,
Peres {\it et al.}~\cite{SRQIT1} have observed that the reduced
spin density matrix of a single spin $1/2$ particle is not a relativistic
invariant, given that Wigner rotations~\cite{Wigner} entangle the
spin with the particle momentum distribution when observed in a
moving referential. This astonishing result, intrinsic and
unavoidable, shows that entanglement theory must be reconsidered
from a relativistic point of view~\cite{BartlettTerno}. On the
other hand, the fundamental implications of relativity on quantum
mechanics could be stronger than what is commonly believed. For
example, Wigner rotations induce also decoherence on two entangled
spins~\cite{PachosSolano,AlsingMilburn,GingrichAdami}. However,
they have not been studied yet in the context of mixed states and
distillable entanglement~\cite{Peres,Horodeckis1}.

A typical situation in SR pertains to a couple of observers: one
is stationary in an inertial frame $\cal{S}$ and the other is also
stationary in an inertial frame $\cal{S}'$ that moves with
velocity $\vec{v}$ with respect to $\cal{S}$. The problems
addressed in SR consider the relation between different
measurements of physical properties, like velocities, time
intervals, and space intervals, of objects as seen by observers in
$\cal{S}$ and $\cal{S}'$. However, in QIT, it is assumed that the
measurements always take place in a proper reference frame, either
$\cal{S}$ or $\cal{S}'$. To see the effects of SR on QIT~\cite{SRQIT}, we need to enlarge the typical
situations where quantum descriptions and measurements take place.

In order to analyze the new possibilities that SR offers, we
introduce the following concepts

\textit{i) Weak isoentangled state} $\rho^{\rm WIE}$: A state that
is entangled in all considered reference frames. This property is
independent of the chosen entanglement measure ${\cal E}$.

\textit{ii) Strong isoentangled state} $\rho^{\rm SIE}_{\cal E}$:
A state that is entangled in all considered reference frames,
while having a constant value associated with a given entanglement
measure ${\cal E}$. This concept depends on the ${\cal E}$ chosen.

\textit{iii) Weak isodistillable state} $\rho^{\rm WID}$: A state
that is distillable in all considered reference frames. This
implies that the state is entangled for these observers.

\textit{iv) Strong isodistillable state} $\rho^{\rm SID}_{\cal
E}$: A state that is distillable in all considered reference
frames, while having a constant value associated with a given
entanglement measure ${\cal E}$. This concept depends on the
${\cal E}$ chosen.

\begin{figure}
\begin{center}
\includegraphics[height=4 cm]{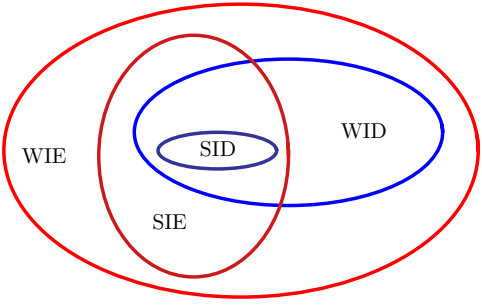}
\end{center}
\caption{(Color online) Hierarchy for the sets of states WIE, SIE,
WID, and SID.\label{grafreldist2}}
\end{figure}

In general, the following hierarchy of sets holds (see Fig.
\ref{grafreldist2} for a pictorial representation)
\begin{equation}
\{\rho^{\rm WIE}\}\supset\{\rho^{\rm SIE}_{\cal
E}\}\supset\{\rho^{\rm SID}_{\cal E}\}\subset\{\rho^{\rm
WID}\}\subset\{\rho^{\rm WIE}\} .
\end{equation}

To illustrate the relative character of distillability, let us
consider the specific situation in which Alice (A) and Bob (B)
share a bipartite mixed state of Werner type with respect to an
inertial frame $\cal{S}$. Moreover, in order to complete the
SR+QIT scenario, we also consider another inertial frame
$\cal{S}'$, where relatives A' and B' of A and B are moving with
relative velocity $\vec{v}$ with respect to $\cal{S}$. Using the
picture of Einstein's trains, we may think that A and B are at the
station platform sharing a set of mixed states, while their
relatives A' and B' are travelling in a train sharing another
couple of entangled particles of the same characteristics. The
mixed state is made up of two particles, say electrons with mass
$m$, having two types of degrees of freedom: momentum $\vec{p}$
and spin $s=\half$. The former is a continuous variable while the
latter is a discrete one. By {\it definition}, we consider our
logical or computational qubit to be the spin degree of freedom.
Each particle is assumed to be localized, as in a box, and its
momentum $\vec{p}$ will be described by the same Gaussian
distribution. We assume that the spin degrees of freedom of
particles $A$ and $B$ are decoupled from their respective momentum
distributions and form the state
\begin{eqnarray}
\rho^{AB}_{\cal{S}}:= F | \Psi^{-}_{\vec{q}} \rangle \langle
\Psi^{-}_{\vec{q}} | && \!\!\!\!\!\! + \frac{1-F}{3} \bigg( |
\Psi^{+}_{\vec{q}} \rangle \langle \Psi^{+}_{\vec{q}} | \nonumber
\\ && + | \Phi^{-}_{\vec{q}} \rangle \langle \Phi^{-}_{\vec{q}} |
+ | \Phi^{+}_{\vec{q}} \rangle \langle \Phi^{+}_{\vec{q}} | \bigg)
. \label{Wernerrela1}
\end{eqnarray}
Here, $F$ is a parameter such that $0 \leq F \leq 1$,
\begin{eqnarray}
\!\!\!\!\!\!\!\! | \Psi^{\pm}_{\vec{q}} \rangle \!\! & := & \!\! 
\frac{1}{\sqrt{2}}\lbrack \Psi_1^{(\rm a)} (\vec{q_a})
\Psi_2^{(\rm b)} (\vec{q_b}) \pm\Psi_2^{(\rm a)}
(\vec{q_a}) \Psi_1^{(\rm b)} (\vec{q_b})\rbrack  , \nonumber \\
\!\!\!\!\!\! | \Phi^{\pm}_{\vec{q}} \rangle \!\!\! &  := & \!\!\!
\frac{1}{\sqrt{2}} \lbrack \Psi_1^{(\rm a)} (\vec{q_a})
\Psi_1^{(\rm b)} (\vec{q_b}) \pm\Psi_2^{(\rm a)}
(\vec{q_a}) \Psi_2^{(\rm b)} (\vec{q_b})\rbrack ,
\label{Bellrela}
\end{eqnarray}
where $\vec{q_a}$ and $\vec{q_b}$ are the corresponding momentum
vectors of particles $A$ and $B$, as seen in $\cal{S}$, and
\begin{eqnarray}
\Psi_1^{(\rm a)} (\vec{q_a}) &:=&  {\cal G} ( \vec{q_a} )
|\!\!\uparrow \rangle = \left(
\begin{array}{cccc}
{\cal G} ( \vec{q_a} ) \\
0 \\
\end{array}
\right) \nonumber \\
\Psi_2^{(\rm a)} (\vec{q_a}) &:=&  {\cal G} ( \vec{q_a} )
|\!\!\downarrow \rangle = \left(
\begin{array}{cccc}
0 \\
{\cal G} ( \vec{q_a} ) \\
\end{array}
\right) \nonumber \\
\Psi_1^{(\rm b)} (\vec{q_b})  &:=&  {\cal G} ( \vec{q_b} )
|\!\!\uparrow \rangle = \left(
\begin{array}{cccc}
{\cal G} ( \vec{q_b} ) \\
0 \\
\end{array}
\right) \nonumber \\
\Psi_2^{(\rm b)} (\vec{q_b})  &:=&  {\cal G} ( \vec{q_b} )
|\!\!\downarrow \rangle = \left(
\begin{array}{cccc}
0 \\
{\cal G} ( \vec{q_b} ) \\
\end{array}
\right) , \label{Skets}
\end{eqnarray}
with Gaussian momentum distributions ${\cal G} ( \vec{q} ):=
\pi^{-3/4}w^{-3/2} \exp ( - {\rm q}^2 /2 w^2 )$, being ${ \rm q}
:= |\vec{q}|$. $|\!\!\uparrow \rangle$ and $|\!\!\downarrow
\rangle$ represent spin vectors pointing up and down along the
$z$-axis, respectively. If we trace momentum degrees of freedom in
Eq.~(\ref{Bellrela}), we obtain the usual spin Bell states, $\{ |
\Psi^{-} \rangle , | \Psi^{+} \rangle, | \Phi^{-} \rangle, |
\Phi^{+} \rangle \}$. If we do the same in
Eq.~(\ref{Wernerrela1}), we remain with the usual spin Werner
state~\cite{Werner}
\begin{eqnarray}
\left(
\begin{array}{cccc}
\frac{1-F}{3} & 0 & 0 & 0 \\
0 & \frac{2F+1}{6} & \frac{1-4F}{6} & 0 \\
0 & \frac{1-4F}{6} & \frac{2F+1}{6} & 0 \\
0 & 0 & 0 & \frac{1-F}{3} \\
\end{array}
\right)  , \label{Werner1}
\end{eqnarray}
written in matrix form, out of which Bell state $| \Psi^{-}
\rangle$ can be distilled if, and only if, $F
> 1 / 2$.

We consider also another pair of similar particles, $A'$ and $B'$,
with the same state as $A$ and $B$, $\rho^{A' B'}_{\cal{S}'} =
\rho^{A B}_{\cal{S}}$, but seen in another reference frame
$\cal{S}'$. The frame $\cal{S}'$ moves with velocity $\vec{v}$
along the $x$-axis with respect to the frame $\cal{S}$. When we
want to describe the state of $A'$ and $B'$ as observed from frame
$\cal{S}$, rotations on the spin variables, conditioned to the
value of the momentum of each particle, have to be introduced.
These conditional spin rotations, considered first by
Wigner~\cite{Wigner}, are a natural consequence of Lorentz
transformations. In general, Wigner rotations entangle spin and
momentum degrees of freedom for each particle. We want to encode
quantum information in the two qubits determined by the spin
degrees of freedom of our two spin-$1/2$ systems. However, the
reduced two-spin state, after a Lorentz transformation, increases
its entropy and reduces its initial degree of entanglement. If we
consider the velocities of the particles as having only non-zero
components in the $z$-axis, each state vector of $A'$ and $B'$ in
Eq.~(\ref{Skets}) transforms as
\begin{eqnarray}
&&\Psi_1(\vec{q})= \left(
\begin{array}{cccc}
{\cal G} ( \vec{q} ) \\
0 \\
\end{array}
\right) \rightarrow \Lambda[\Psi_1(\vec{q})]=\left(
\begin{array}{cccc}
\cos \theta_{\vec{q}} \\
\sin \theta_{\vec{q}} \\
\end{array}
\right) {\cal G} ( \vec{q} )
\nonumber \\
&& \Psi_2(\vec{q})=\left(
\begin{array}{cccc}
0 \\
{\cal G} ( \vec{q} ) \\
\end{array}
\right) \rightarrow \Lambda[\Psi_2(\vec{q})]=\left(
\begin{array}{cccc}
-\sin \theta_{\vec{q}} \\
\,\,\,\,\, \cos \theta_{\vec{q}} \\
\end{array}
\right) {\cal G} ( \vec{q} ) ,\nonumber\\ \label{transfrules}
\end{eqnarray}
where $\cos \theta_{\vec{q}}$ and $\sin \theta_{\vec{q}}$ express
Wigner rotations conditioned to the value of the momentum vector.

The most general bipartite density matrix in the rest frame for
arbitrary spin-1/2 states and Gaussian product states in momentum,
is spanned by the tensor products of $\Psi^{(\rm a)}_1$,
$\Psi^{(\rm a)}_2$, $\Psi^{(\rm b)}_1$, and $\Psi^{(\rm b)}_2$,
and can be expressed as
\begin{equation}
\rho=\sum_{ijkl=1,2}C_{ijkl}\Psi^{(\rm a)}_i( \vec{q_a}
)\otimes\Psi^{(\rm b)}_j( \vec{q_b} ) [\Psi^{(\rm a)}_k( \vec{q_a'}
)\otimes\Psi^{(\rm b)}_l( \vec{q_b'}
)]^{\dag}.\label{densitygeneral}
\end{equation}
Under a boost, Eq. (\ref{densitygeneral}) will transform into
\begin{eqnarray}
\Lambda\rho\Lambda^{\dag}\!\!\! & = &\!\!\!\!\!
\sum_{ijkl=1,2}C_{ijkl}\Lambda^{(\rm a)}[\Psi^{(\rm a)}_i(
\vec{q_a} )]\otimes
\Lambda^{(\rm b)}[\Psi^{(\rm b)}_j( \vec{q_b} )]\nonumber\\
&&\times\{\Lambda^{(\rm a)}[\Psi^{(\rm a)}_k( \vec{q_a'}
)]\otimes\Lambda^{(\rm b)}[\Psi^{(\rm b)}_l( \vec{q_b'}
)]\}^{\dag}.\label{densityboost}
\end{eqnarray}
Tracing out the momentum degrees of freedom, we obtain
\begin{eqnarray}
& \mathrm{Tr}_{\rm \vec{q_a}, \vec{q_b}} &
(\Lambda\rho\Lambda^{\dag})\nonumber \\  & = & \!\!\!\!\!\!\!\!
\sum_{ijkl=1,2}C_{ijkl}\mathrm{Tr}_{\rm \vec{q_a}}(\Lambda^{(\rm
a)}[\Psi^{(\rm a)}_i( \vec{q_a} )]\{\Lambda^{(\rm a)}[\Psi^{(\rm
a)}_k( \vec{q_a})]\}^{\dag}) \nonumber \\ && \otimes
\mathrm{Tr}_{\rm \vec{q_b}}(\Lambda^{(\rm b)}[\Psi^{(\rm b)}_j(
\vec{q_b} )]\{\Lambda^{(\rm b)}[\Psi^{(\rm b)}_l(
\vec{q_b})]\}^{\dag}).\label{densityboosttr}
\end{eqnarray}
Following Peres {\it et al.}~\cite{SRQIT1}, we compute the Lorentz
transformed density matrix of state $\Psi_1$, after tracing out
the momentum. The expression, to first order in $w/m$, reads
\begin{equation}
\mathrm{Tr}_{\rm
\vec{q}}[\Lambda\Psi_1(\Lambda\Psi_1)^{\dag}]=\frac{1}{2}
\left(\begin{array}{cc}1+n_z' & 0\\0 &
1-n_z'\end{array}\right),\label{peresmatriz1}
\end{equation}
where $n'_z:=1-\left(\frac{w}{2m}\tanh \frac{\alpha}{2}\right)^2$
and $\cosh\alpha:=\gamma=(1-\beta^2)^{-1/2}$. Larger values of $w / m$ are possible and mathematically correct~\cite{GingrichAdami}, though not necessarily physically consistent. First, the Newton-Wigner localization
problem~\cite{Sakurai} prevents us from considering momentum
distributions with $w \lesssim m$. In that case, particle creation
would manifest and our model, relying on a bipartite state of the
Fock space, would break down. Second, $w \sim m$ would produce fast wave-packet spreading, yielding an
undesired particle delocalization.

\begin{figure}
\begin{center}
\hspace*{-0.6cm}
\includegraphics[height=4.5cm]{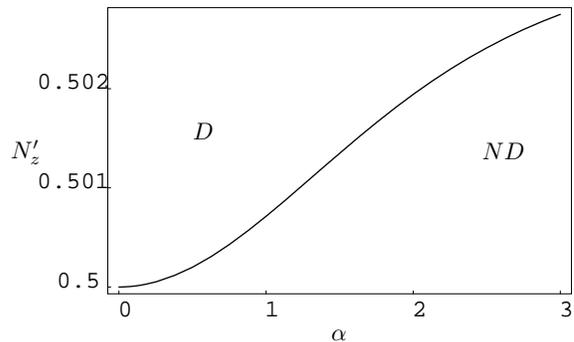}
\end{center}
\caption{$N'_z$ of Eq.~(\ref{eqNz}) vs. the rapidity $\alpha$, for
$w/2m=0.1$. \label{grafreldist}}
\end{figure}

This can be generalized to the other three tensor products
involving $\Psi_1$ and $\Psi_2$,
\begin{eqnarray}
\mathrm{Tr}_{\rm \vec{q}}[\Lambda\Psi_2(\Lambda\Psi_2)^{\dag}]& =
&\frac{1}{2}\left(\begin{array}{cc}1-n_z'
& 0\\0 & 1+n_z'\end{array}\right),\label{peresmatriz2}\\
\mathrm{Tr}_{\rm
\vec{q}}[\Lambda\Psi_1(\Lambda\Psi_2)^{\dag}]&=
&\frac{1}{2}\left(\begin{array}{cc}0
& 1+n'_z\\-(1-n_z') & 0\end{array}\right),\label{peresmatriz3}\\
\mathrm{Tr}_{\rm
\vec{q}}[\Lambda\Psi_2(\Lambda\Psi_1)^{\dag}]&=&\frac{1}{2}
\left(\begin{array}{cc}0
& -(1-n'_z)\\1+n_z' & 0\end{array}\right).\label{peresmatriz4}
\end{eqnarray}
With the help of Eqs.~(\ref{densityboosttr}-\ref{peresmatriz4}),
it is possible to compute the effects of the Lorentz
transformation, associated with a boost in the $x$-direction, on
any density matrix of two spin-1/2 particles with factorized
Gaussian momentum distributions. In particular,
Eq.~(\ref{Wernerrela1}) is reduced to
\begin{equation}
 \left( \!\!\!
\begin{array}{cccc}
\frac{1}{4} + c_F {n'_z}^2 & 0 & 0 &
c_F ({n'_z}^2-1) \\
0 & \frac{1}{4} - c_F {n'_z}^2 &
c_F({n'_z}^2+1) & 0 \\
0 & c_F ({n'_z}^2+1) & \frac{1}{4}
- c_F {n'_z}^2 & 0 \\
c_F ({n'_z}^2-1) & 0 & 0 &
\frac{1}{4} + c_F {n'_z}^2 \\
\end{array}
\!\!\! \right) , \\ \label{Werner2}
\end{equation}
where $c_F:=\frac{1-4F}{12}$.  We can apply now the positive
partial transpose (PPT) criterion~\cite{Peres,Horodeckis1} to know
whether this state is entangled and distillable. Due to the
box-inside-box structure of Eq.~(\ref{Werner2}), it is possible to
diagonalize its partial transpose in a simple way, finding the
eigenvalues
\begin{eqnarray}
x_1 = \frac{2F+1}{6} \,\, , \hspace*{0.5cm}
x_2 = \frac{1-F}{3} + \frac{1-4F}{6} {n'_z}^2 \,\, , \nonumber \\ [0.1cm]
x_3 = \frac{1-F}{3} -\frac{1-4F}{6} {n'_z}^2 \,\, , \hspace*{0.5cm}
x_4 = \frac{2F+1}{6} \,\, .
\end{eqnarray}
Given that $F > 0$, $x_1$ and $x_4$ are always positive, and also $x_3$ for $0 < {n'_z} < 1$. The eigenvalue
$x_2$ is negative if, and only if, $F
> N'_z$, where $N'_z:=(2 + {n'_z}^2)/(2 + 4{n'_z}^2)$. The latter
implies that in the interval
\begin{eqnarray}
\frac{1}{2} < F < N'_z \label{eqNz}
\end{eqnarray}
distillability of state $| \Psi^{-} \rangle$ is possible for the
spin state in A and B, but impossible for the spin state in A' and
B', both described in frame $\cal{S}$. We plot in
Fig.~\ref{grafreldist} the behavior of $N'_z$ as a function of the
rapidity $\alpha$. The region below the curve (ND) corresponds to
the $F$ values for which distillation is not possible in the Lorentz
transformed frame. On the other hand, the region above the curve
(D), corresponds to states which are distillable for the
corresponding values of $n'_z$. Notice that there are values of $F$
for which the Werner states are weak isodistillable and weak
isoentangled, corresponding to the states in the region D above the
curve for the considered range of $n'_z$. On the other hand, there
are states that will change from distillable (entangled) into
separable for a certain value of $n'_z$, showing the relativity of
distillability and separability.

The study of strongly isoentangled and strongly isodistillable two-spin
states is a much harder task that will depend on the entanglement measure we choose. We believe that these cases impose demanding
conditions and, probably, this kind of states does not exist. However we would like to give a plausibility argument
to justify this conjecture. Our
argument is based on two mathematical points: (i) analytic continuation is a
mathematical tool that allows to extend the analytic behavior of a
function to a region where it was not initially defined, and (ii) an
analytic function is either constant or it changes along all its interval of definition. Point (i) will allow us to extend analytically our calculation to $n'_z=0$, an unphysical but mathematically convenient limit. Point (ii) will be applied to any well-behaved entanglement measure. We
consider then a general spin density matrix
\begin{eqnarray}
 \hspace*{-0.6cm}
\rho:=\left( \!\!\!
\begin{array}{cccc}
a_1 & b_1 & b_2  &
b_3  \\
b_1^* & a_2 & c_1 & c_2 \\
b_2^* & c_1^* & a_3 & d\\
b_3^* & c_2^* & d^* & a_4 \\
\end{array}
\!\!\! \right), \label{generaldensity}
\end{eqnarray}
where $a_1$, $a_2$, $a_3$, and $a_4$ are real, and $\sum_i a_i=1$.
The analytic continuation of the Lorentz transformed state,
according to Eqs. (\ref{densityboosttr}-\ref{peresmatriz4}), in
the limit $n'_z\rightarrow 0$, is
\begin{eqnarray}
\hspace*{-0.6cm} \left( \!\!\!
\begin{array}{cccc}
1/4 & \frac{i(\mathrm{\Im} b_1+\mathrm{\Im} d)}{2} &
\frac{i(\mathrm{\Im} b_2+\mathrm{\Im} c_2)}{2}  &
\frac{(\mathrm{\Re} b_3-\mathrm{\Re} c_1)}{2}   \\
\frac{-i(\mathrm{\Im} b_1+\mathrm{\Im} d)}{2} & 1/4 &
\frac{(-\mathrm{\Re} b_3+\mathrm{\Re} c_1)}{2} &
\frac{i(\mathrm{\Im} b_2+\mathrm{\Im} c_2)}{2} \\
\frac{-i(\mathrm{\Im} b_2+\mathrm{\Im} c_2)}{2} &
\frac{(-\mathrm{\Re} b_3+\mathrm{\Re} c_1)}{2} & 1/4 &
\frac{i(\mathrm{\Im} b_1+\mathrm{\Im} d)}{2}\\ \frac{(\mathrm{\Re}
b_3-\mathrm{\Re} c_1)}{2} & \frac{-i(\mathrm{\Im} b_2+\mathrm{\Im}
c_2)}{2} & \frac{-i(\mathrm{\Im} b_1+\mathrm{\Im} d)}{2} & 1/4 \\
\end{array}
\!\!\! \right),  \label{separable}
\end{eqnarray}
where $\Re$ and $\Im$ denote the real and imaginary parts.
 This state is separable because its eigenvalues, given by
\begin{eqnarray}
\lambda_{1,2}&=&\frac{1}{4}[1-2
\mathrm{\Re}(b_3-c_1)\pm 2\mathrm{\Im}(b_1+b_2+c_2+d)]\nonumber\\
\lambda_{3,4}&=&\frac{1}{4}[1+2
\mathrm{\Re}(b_3-c_1)\pm 2\mathrm{\Im}(b_1-b_2-c_2+d)]\nonumber\\
\,\,\,\,\,\,\,\,\,\, \label{autovisoentang}
\end{eqnarray}
coincide with the corresponding ones for the partial transpose
matrix. In this case, $\lambda_1\leftrightarrow\lambda_4$, and
$\lambda_2\leftrightarrow\lambda_3$. So, according to the PPT
criterion, the analytic continuation of the Lorentz transformed
density matrix of all two spin-1/2 states, with factorized
Gaussian momentum distributions, converges to a separable state in
the limit of $n'_z\rightarrow 0$~\cite{comment}. Our analytic
calculation holds for $n'_z \lesssim 1$, leaving out of reach the
case $n'_z=0$. However, any analytic measure of entanglement, due
to this behavior of the analytic continuation at $n'_z=0$, is
forced to change with $n'_z$ for $n'_z \lesssim 1$, except for
states separable in all frames. In this way, we give evidence of the non-existence of strong
isoentangled and isodistillable states, for variations of
the parameter $n'_z$ under the present assumptions.

From a broader perspective, our analysis considered the invariance
of entanglement and distillability of a two spin-$1/2$ system
under a particular completely positive (CP) map, the one
determined by the local Lorentz-Wigner transformations. The study
of similar properties in the context of general CP maps is an
important problem that, to our knowledge, has not received much
attention in QIT, and that will require a separate and more
abstract analysis. Moreover, for higher dimensional spaces, like a
two spin-$1$ system (qutrits), the notion of relativity of bound
entanglement will also arise~\cite{Horodeckis2}.

In summary, the concepts of weak and strong isoentantangled and
isodistillable states were introduced, which should help to
understand the relationship between special relativity and quantum
information theory. The study of Werner states allowed us to show
that distillability is a relative concept, depending on the frame
in which it is observed. We have proven the existence of weak
isoentangled and weak isodistillable states in our range of
validity of the parameter $n'_z$. We also conjectured the
non-existence of strong isoentangled and isodistillable
two-spin states. We give evidence for this
result relying on the analytic continuation of the
Lorentz transformed spin density matrix for a general two spin-1/2
particle state with factorized momentum distributions.

L.L. acknowledges financial support from Spanish MEC through FPU
grant AP2003-0014, CSIC 2004 5 0E 271 and FIS2005-05304  projects. M.A.M.-D. thanks DGS for grant under contract BFM 2003-05316-C02-01. E.S. acknowledges support
from SFB 631, EU RESQ and EuroSQIP projects.


\begin{thebibliography}{99}
\bibitem{comment} We consider $n'_z \rightarrow 0$ to support our conjecture, though
we know from previous arguments that it is unphysical.
\end{thebibliography}
\end{document}